\documentclass[10pt]{article}
\usepackage{amsmath, amssymb}
\usepackage[mathscr]{eucal}

\allowdisplaybreaks

\usepackage[paper=a4paper,
            lmargin=3.0cm,rmargin=2.9cm,tmargin=3.8cm,bmargin=3.0cm,%
            foot=0.9truecm]{geometry}
\newfont{\bssfont}{cmssbx10}

\def\unomasuno{{1\!+\!1}}
\def\smallonehalf{\frac{{}_1}{{}^2}}

\def\={\!=\!}
\def\>{\!>\!}
\def\<{\!<\!}

\def\k{\kappa}
\def\ki{\kappa_1}
\def\kii{\kappa_2}

\def\quadrant{{\vert\kern-.15em\underline{\, \cdot\, } \,} } 

\def\CKspace{{\textbf{\itshape S}}^{\,2}_{\ki\![\kii]}}
\def\CKspaceKepMom{{\textbf{\itshape S}}^{\,2}_{\sLab\![\kii]}}

\def\CKC{\,\mathop{\kern-.15em\rm C}\nolimits}    
\def\CKS{\,\mathop{\kern-.15em\rm S}\nolimits}    
\def\CKT{\,\mathop{\kern-.15em\rm T}\nolimits}    
\def\CKV{\,\mathop{\kern-.15em\rm V}\nolimits}    
\def\CKtoV{\,\mathop{\kern-.15em\rm W}\nolimits}    

\def\CKArcT{\,\mathop{\kern-.15em\rm ArcT}\nolimits}    

\def\CKc{\CKC_{\ki}\!}
\def\CKs{\CKS_{\ki}\!}
\def\CKt{\CKT_{\ki}\!}
\def\CKv{\CKV_{\ki}\!}
\def\CKcc{\CKC_{\kii}\!}
\def\CKss{\CKS_{\kii}\!}
\def\CKtt{\CKT_{\kii}\!}
\def\CKvv{\CKV_{\kii}\!}
\def\CKccc{\CKC_{\ki\!\kii}\!}
\def\CKsss{\CKS_{\ki\!\kii}\!}
\def\CKttt{\CKT_{\ki\!\kii}\!}

\def\CKcSq{\CKC^2_{\ki}\!}  
\def\CKsSq{\CKS^2_{\ki}\!}  \def\CKsCu{\CKS^3_{\ki}\!}
\def\CKtSq{\CKT^2_{\ki}\!}

\def\CKcccSq{\CKC^2_{\ki\!\kii}\!}

\def\CKCr{\CKc(r)}  \def\CKCrSq{\CKcSq(r)}  
\def\CKCu{\CKc(u)}  
\def\CKCx{\CKc(x)}  \def\CKCxSq{\CKcSq(x)}
\def\CKSr{\CKs(r)}  \def\CKSrSq{\CKsSq(r)}  
\def\CKSu{\CKs(u)}
\def\CKSx{\CKs(x)}
\def\CKTr{\CKt(r)}  \def\CKTrSq{\CKtSq(r)}
\def\CKTu{\CKt(u)}

\def\CKCphi{\CKcc(\phi)}  
\def\CKSphi{\CKss(\phi)}  
\def\CKTphi{\CKtt(\phi)}  
\def\CKVphi{\CKvv(\phi)}

\def\CKCy{\CKccc(y)} \def\CKCySq{\CKcccSq(y)}
\def\CKCv{\CKccc(v)} 
\def\CKSy{\CKsss(y)} 
\def\CKSv{\CKsss(v)}
 
\def\CKTv{\CKttt(v)} 

\def\CKP{\mathcal{P}}
\def\CKJ{\mathcal{J}}
\def\CKW{\mathcal{W}}

\def\CKF{\mathcal{F}}

\def\CKLag{\mathcal{L}}
\def\CKPot{\mathcal{V}}
\def\CKE{\mathcal{E}}

\def\cu{C_{\ki}\!}  

\def\IL{\relax{\rm I\kern-.18 em L}}

\def\sLab{\sigma}
\def\CKcLab{\CKC_{\sLab}\!}
\def\CKsLab{\CKS_{\sLab}\!}
\def\CKtLab{\CKT_{\sLab}\!}

\begin{document}


\title{Levi-Civita regularization and geodesic flows for the `curved' Kepler problem}

\author{Leonor Garc\'{\i}a-Guti\'errez \footnote{\sl E-mail address:  %
       {leonor.garcia@fta.uva.es}}
        and
        Mariano Santander \,\footnote{\sl E-mail address: {msn@fta.uva.es}} \\[4pt]
{\enskip}
   {\sl Departamento de F\'{\i}sica Te\'orica, Facultad de Ciencias}\\
   {\sl Universidad de Valladolid,  47011 Valladolid, Spain} }
\maketitle

\begin{abstract}
We introduce the regularization Levi-Civita parameter  for the `curved Kepler', i.e., motion under the `Kepler-Coulomb' potential in a configuration space with any constant curvature and metric of any signature type. Consistent use of this  parameter allows to solve the problem of motion (orbit shape and time evolution along the orbit), thereby extending the use of the Levi-Civita parameter beyond the usual Kepler problem in a flat Euclidean configuration space. A `universal' description, where all relations are applicable to the motions in any space and with any energy follow from our approach, with no need to discuss separately the cases where the configuration space is flat or where energy vanishes. 

We also discuss the connection of this `curved Kepler' problem with a geodesic flow. The well known results by Moser, Osipov and Belbruno are shown to hold essentially unchanged beyond the flat Euclidean configuration space. `Curved' Kepler motions with a fixed value of the constant of motion $-(2E - \kappa_1\kappa_2\CKJ^2)$ on any curved configuration space with constant curvature $\kappa_1$ and metric of signature type $\kappa_2$ can be identified with the geodesic flow on a space with curvature $\sLab$ and metric of the same signature type.

\end{abstract}
 







\section{Introduction}

For the Kepler motion in the Euclidean plane, the simplest way to integrate the Newton's equations involves an apparently artificial construct: by replacing the physical time $t$ by a particular {\it fictitious time} $s$, called {\it Levi-Civita regularization parameter}, closed (and rather simple) explicit expressions for the cartesian $x,y$ and polar coordinates $r, \phi$ and, more importantly, for the time $t$ can be given as functions of $s$ (see e.g., Milnor \cite{MilnorGKP}). The dependence in $s$ of all these functions is smooth, so this parameter in addition provide a way to {\it regularize} the description when the orbit approaches a {\it collision} orbit (where position and velocity are linearly dependent vectors). The same parameter $s$ appears also very directly in the relation between the Kepler problem and the geodesic flow on spaces of constant curvature \cite{Moser70, Os72, Os77, Be77, Be81, MilnorGKP, AnosovGKP}, and it turns out to be proportional also to the `classical' parameter for studying Kepler evolution, called the {\it eccentric anomaly} (for elliptic orbits this parameter was introduced by Kepler himself \cite{Ke1609}). Hence the Levi-Civita regularization parameter plays a {\it central} role in the Kepler problem. 

Our aim in this paper is to find the regularization parameter for the `curved' Kepler problem, this is, the analogue of the Kepler problem when the configuration space is no longer the Euclidean space, but has a (non-zero) constant curvature.  We perform this task using a Cayley-Klein type description, which allows to deal in a single run with the Kepler problem in a space $\CKspace$ depending on two real parameters $\ki$ and $\kii$. The constant curvature of this space is $\ki$ and its metric is either Riemannian, degenerate or Lorentzian, corresponding to the three alternatives $\kii\>, \=, \<0$ for the second parameter $\kii$, referred to as the `signature type'. As an added bonus of this approach, a complete and fully explicit solution for the `Kepler problem' in a Lorentzian configuration space (DeSitter or Minkowski) is obtained. 

There are some papers dealing with particular aspects of either the classical or the quantum Kepler problem in configuration spaces of constant curvature but none of them (to our knowledge) deals with the regularization parameter for the problem, nor its possible relation with a geodesic flow. And in spite of recent interest in studying motion in Lorentzian manifolds, we do not know either any paper dealing with the `Kepler' problem on a Lorentzian configuration space, curved or not. 
Thus, when the configuration space has nonzero curvature, or when the metric is of Lorentzian signature, the results obtained are new. 

We solve completely this problem, and recover as a particular instance all the well known results for the `flat' Kepler problem in Euclidean space. In particular, the close relationship between the set of Euclidean Kepler motions with total energy $E$ and the geodesic flow in a space of constant curvature $-2E$ \cite{Moser70, Os72, Os77, Be77, Be81, MilnorGKP}, appears as the particular `flat case' of a {\it generic} relation holding for the `curved' Kepler motions in a configuration space $\CKspace$ of any constant curvature and either metric signature type. If $\sLab$ denotes the combination $\sLab=-(2E-\ki\kii\CKJ^2)$ of energy and angular momentum, the result we obtain is: the set of `curved' Kepler motions with a given constant value of $\sLab$ can be identified to a geodesic flow in a space of constant curvature $\sLab$ whose metric has the same signature type as the configuration space. This result extends directly the Euclidean Kepler one, yet we have not found any reference to this, nor to any similar result, in the literature.   

All the expressions we give are completely explicit; for the Kepler motion, we disclose the dependance of all the relevant coordinates on the `curved' Levi-Civita parameter $s$. This automatically produces a large number of results and relations for the `curved' Kepler problem, which extend properties well known for the Euclidean Kepler problem: the `cycle' character of the Kepler hodographs, the Kepler equation and the dependance $t(s)$, etc. 
The connection of Kepler motions with fixed value of the quantity $\sLab$ with a geodesic flow in an auxiliar space whose curvature is precisely $\sLab$ is quite direct. 

This is done using as a language the parametric CK type approach, which allows to do computations for all the configuration spaces at a single run, considering $\ki, \kii$ as free parameters. But there is more: even if we stick to studying the Kepler problem in, say,  Euclidean space ${\bf E}^2$, where the two CK free parameters have fixed values $\ki\=0, \kii\=1$, the quantity $\sLab$, which is in this case the energy up to a factor, which plays a role as a {\it third Cayley-Klein parameter}; working consistently in these terms allows to give an unified description, for all types of orbits, in a single run, and with a dependence on $\sLab$ which is smooth when $\sLab\to0$. The same happens when the configuration space is the general $\CKspace$. Thus the CK type approach provides some new perspective to an `universal' formulation for the Keplerian orbits, encompassing not only all energies but also all possible values of the curvature and signature type of the configuration space. 

The plan of the paper is the following: Section 2 is devoted to the regularization of the `curved' Kepler problem. First we give the basics on dynamics on a space $\CKspace$ and introduce the `curved' Kepler potential and the `curved' Levi-Civita parameter $s$. Then we derive expressions for coordinates and time in the curved case as functions of $s$. Section 3 is devoted to the connection between Kepler motion in $\CKspace$ and a geodesic flow. This is done directly, in terms of a stereographic projection, which identifies `curved' Kepler motions with a fixed value of the constant of motion $\sLab = -(2E - \ki\kii\CKJ^2)$ to the geodesic flow on a space with curvature $\sLab$ and metric of the same signature type as the configuration space. Again this reduces, when the configuration space is Euclidean, to the well known Moser--Osipov--Belbruno result: Kepler motions in ${\bf E}^2$ with energy $E$ can be identified with the geodesic flow on a Riemannian space with  constant curvature $-2E$. The three `curved' Kepler laws are stated and discussed in Subsection 3.4. Finally, Section 4 
discusses the specialization to the Euclidean configuration space, translating the results to the standard language and relating the parameter $s$ to the eccentric anomaly.  

\section{The regularization of the curved Kepler problem}

\subsection{Dynamics in a configuration space $\CKspace$}

We denote by $\CKspace$ a 2d space with constant curvature $\ki$ and metric of a signature type $(1, \kii)$. By rescaling lenghts and angles, the two parameters $\ki, \kii$ could be brought independently to a {\it standard} value $1, 0, -1$ and the nine combinations correspond to the so called `standard' Cayley-Klein CK spaces. Instead of reducing $\ki, \kii$ some standard values to start with, we will leave both $\ki, \kii$ as free parameters. There are two main reasons to do so. First, a unique computation, only slightly more complicated than the one required for each individual instance, covers all nine cases. Second, dealing with general values for $\ki, \kii$, one can get a valuable perspective on how different properties and relations depend on the curvature and/or signature type of the space, and how things change when curvature vanishes or changes sign, or when the metric degenerates or changes from positive definite to indefinite. This makes a subsequent analysis of contractions and limiting cases completely redundant. For more details on this formalism, see \cite{HeOrSa00, ConformalHS02, ConformalCompactHS02}.

The basic tool in this CK approach is the use of a set of `labeled' trigonometric functions. The $\kappa$-labeled `Cosine' $\CKC_{\k}(x)$ and `Sine' $\CKS_{\k}(x)$ functions are defined here as the solutions of the differential equation:
\begin{equation}
\displaystyle\frac{d^2}{dx^2} F(x) = -\k\, F(x)\,,
\label{DifEqDefSinCos}
\end{equation}
determined respectively by the initial conditions:
\begin{equation}
\CKC_{\k}(0)=1, \quad 
\displaystyle\left.\frac{d \CKC_{\k}(x)}{dx}\right|_{x=0}  = 0\,;
\qquad
\CKS_{\k}(0)=0, \quad 
\displaystyle\left.\frac{d \CKS_{\k}(x)}{dx}\right|_{x=0}  = 1\,.
\end{equation}
In the field of abstract differential equations, a similar approach leads to the `cosine' and `sine' families \cite{Fat}. Here, with the simplest equation (\ref{DifEqDefSinCos}) this `cosine' and `sine' are ordinary functions, which admit the following expressions, with an analytic dependence in both variables $x$ and $\k$:
\begin{equation}
\CKC_{\k}(x) :=\left\{
\begin{array}{l}
  \cos {\sqrt{\k}\, x} \\[2pt]
  1  \\[2pt]
\cosh {\sqrt{-\k}\, x} 
\end{array}\right. ,
\qquad
\CKS_{\k}(x) :=\left\{
\begin{array}{ll}
    \frac{1}{\sqrt{\k}} \sin {\sqrt{\k}\, x} &\qquad  \k >0 \\[2pt]
  x &\qquad  \k  =0 \\[2pt]
\frac{1}{\sqrt{-\k}} \sinh {\sqrt{-\k}\, x} &\qquad   \k <0
\end{array}\right.,
\label{CKDefSinCos}
\end{equation}
These two functions satisfy the basic identity (as well as many others, see e.g. \cite{HeOrSa00}):
\begin{equation}
{\CKC^2_{\k}}(x)  + \k\, {\CKS^2_{\k}}(x) = 1
\label{BasicIdSinCos}
\end{equation}

The `Tangent' $\CKT_{\k}(x)$ is defined as the quotient $\CKT_{\k}(x)= \CKS_{\k}(x)/\CKC_{\k}(x)$. Another function appearing naturally is the $\k$ version of the `versed sine', defined as $\CKV_{\k}(x)=(1-\CKC_{\k}(x))/\k$; it is interesting to realize that when $\k\to0$, both numerator and denominator tend to $0$ in such a way that $\CKV_{\k}(x)$ is well defined even when $\k\to0$ and $\CKV_{0}(x)=x^2/2$. The inverse functions are denoted accordingly; here only $\CKArcT_\k(x)$ will appear. 

These functions allow us to write expressions in any $\CKspace$ in a  unified way. For the two standard choices $\k=\pm 1$, these functions are precisely the circular or hyperbolic trigonometric functions. The singular case $\k=0$ corresponds to the so-called parabolic trigonometric functions, and for general $\k$, the  labeled `cosine' $\CKC_{\k}(x)$, `sine' $\CKS_{\k}(x)$ and `versed sine' $\CKV_{\k}(x)$ functions can be considered as an one-parameter set of deformations of the corresponding `parabolic' functions, equal respectively to $1, x$ and $x^2/2$, to which these reduce for $\k=0$. Within this viewpoint, the tangent $\CKT_{\k}(x)$ is to be considered as another different deformation of the function $x$.

\begin{table}[h]
{\footnotesize
 \noindent
\caption{{The nine standard two-dimensional CK
spaces $\CKspace$.}}\label{table:9CKGeometries}
\noindent\hfill
\begin{tabular}{llll}
\hline\\[-8pt]
 &\multicolumn{3}{c}{Measure of distance \& Sign of $\ki$}\\[2pt]
\cline{2-4}\\[-8pt]
Measure of angle\ &Elliptic&Parabolic&Hyperbolic\\
\ \ \& Sign of $\kii$&$\ki=1$&$\ki=0$&$\ki=-1$\\[2pt]
\hline\\[-8pt]\hline\\[-8pt]
&Elliptic&Euclidean&Hyperbolic\\
Elliptic $\kii=1$&${\bf S}^2$&${\bf E}^2$&${\bf H}^2$\\[2pt]
\hline\\[-8pt]
&Co-Euclidean&Galilean&Co-Minkowskian\\
&Oscillating NH & &Expanding NH\\
Parabolic $\kii=0$&${\bf ANH}^\unomasuno$&${\bf G}^\unomasuno$&${\bf NH}^\unomasuno$\\[2pt]
\hline\\[-8pt]
&Co-Hyperbolic&Minkowskian&Doubly Hyperbolic\\
&Anti-de Sitter& &De Sitter\\
Hyperbolic $\kii=-1$&${\bf AdS}^\unomasuno$&${\bf M}^\unomasuno$&${\bf dS}^\unomasuno$\\[2pt]
\hline\\[-8pt]\hline\\[-8pt]
\end{tabular}\hfill}
\end{table}

The space $\CKspace$ involves {\it two} independent labels $\ki, \kii$, and in its CK description $\k$-labeled functions with the labels $\ki, \ \kii, \ \ki\kii$ appear. When the two basic labels are positive, the space $\CKspace$ is a two-dimensional sphere; the {\it standard} ${\bf S}^2$ corresponds to the choice $\ki\=1,\, \kii\=1$; other {\it standard} choices are $\ki\=0, \kii\=1\ (\equiv{\bf E}^2$, the Euclidean plane) or $\ki\=-1, \kii\=1\ (\equiv{\bf H}^2$, the hyperbolic or Lobachewski plane). The remaining standard spaces are the  three `non-relativistic' (with absolute time) space-times, appearing for $\kii=0$:  antiNewton-Hooke space ${\bf ANH}^{\unomasuno}$, Galilean space ${\bf G}^{\unomasuno}$, Newton-Hooke space ${\bf NH}^{\unomasuno}$, and the three relativistic space-times in $\unomasuno$ dimensions, appearing for $\kii<0$: the AntiDeSitter sphere ${\bf AdS}^{\unomasuno}$, Minkowskian space ${\bf M}^{\unomasuno}$ and deSitter sphere ${\bf dS}^{\unomasuno}$). This information is displayed in the Table; see \cite{ConformalHS02, ConformalCompactHS02} for more comments.

At a first contact with this formalism, the reader might well pretend that $\ki=1, \kii=1$, and all the labeled `cosine' $\CKC_{\k}(x)$ and `sine' functions $\CKS_{\k}(x)$ with either  label ($\ki, \kii$ or $\ki\kii$, all equal to 1) are ordinary, circular cosines $\cos(x)$ and sines $\sin(x)$. With this understanding, every relation will apply to the standard sphere ${\bf S}^2$, where it is possible to visualize most properties in an easy way. At the final {\it intended level} of reading, $\ki, \kii$ should be considered of course as free parameters, and the CK formalism keeps track automatically of all sign differences, vanishing of some terms, replacement of (some) circular trigonometric functions by their parabolic or hyperbolic analogues, etc., which distinguish relations in the nine CK spaces. 

As a first example of this language, let us write the expressions for the metric in the $\CKspace$, in the intrinsic polar $(r, \phi)$, parallel `1' $(x, v)$ and parallel `2' $(u,y)$ coordinates:
\begin{equation}
dl^2 = 
   dr^2 + \kii\CKSrSq \,d \phi ^2=
      dx^2 + \kii\CKCxSq \,dv^2 = 
   \CKCySq du^2 + \kii\, dy^2.
\end{equation}
Comparison with the very well known form for the metrics in some standard spaces 
as ${\bf S}^2, {\bf E}^2, {\bf H}^2$ or ${\bf M}^{\unomasuno}$ will give some feeling on the unification capacity of the CK formalism. 
In the general curved CK space the quantities $u$ and $x$ (or $v$ and $y$) are generally different (this is clear on a sphere); this follows from the relations holding for the general CK space $\CKspace$:
\begin{equation}
\CKTu = \CKTr \CKCphi, \qquad \CKTv = \CKTr \CKSphi
\label{RelsCoordUV_RPhi}
\end{equation}
\begin{equation}
\CKSx = \CKSr \CKCphi, \qquad \  \CKSy = \CKSr \CKSphi
\label{RelsCoordXY_RPhi}
\end{equation}
which can be considered as formulas for orthogonal triangles \cite{HeOrSa00} in the trigonometry of $\CKspace$. When $\ki=0$ these relations imply a characteristic degeneracy of the {\it flat} spaces: the equalities $u = x$ and $v = y$. From the point of view of deformations, (\ref{RelsCoordUV_RPhi}) and (\ref{RelsCoordXY_RPhi}) can be looked at  as two possible `curved' generalizations of the Euclidean relations $x=r \cos\phi,\  y = r \sin\phi$. Notice that $(u, y)$ and $(x, v)$ are {\it orthogonal} coordinates in any $\CKspace$, but $(u, v)$ or $(x, y)$ are orthogonal systems only when the curvature $\ki$ vanishes; in this case all these four systems collapse to a single Cartesian system. 

Consider now the motion of a particle in the configuration space $\CKspace$ under a natural mechanical type Lagrangian, with a kinetic term given by the metric and possibly a potential depending on the coordinates:  
\begin{equation}
\CKLag= \smallonehalf\, g_{\mu\nu}(q^1, q^2) \dot{q^\mu}\dot{q^\nu} - \CKPot(q^1, q^2).  
\end{equation} 
Constants of motion for this Lagrangian which are {\it linear in the velocities} occur only if the potential is invariant under some one-parameter subgroup of isometries, and are equal to the corresponding Noether momenta. With a scale factor suitable to simultaneously deal with all the spaces in the CK family \cite{CRS07a, CRS07c}, the natural base of three Noether momenta $\CKP_1, \CKP_2, \CKJ$ is given in the three coordinates as: 
\begin{equation}
\left( \!\!\begin{tabular}{c} 
$\CKP_1$ \cr $\CKP_2$ \cr $\CKJ$ 
\end{tabular} \!\!\! \right)\!\!
=
\!\!\left(\!\!\begin{tabular}{c} 
$\CKCphi \dot{r} - \kii\CKCr \CKSr \CKSphi\dot{\phi}$\cr 
$\CKSphi\dot{r} + \CKCr \CKSr \CKCphi\dot{\phi}$\cr 
$\CKSrSq\dot{\phi}$
\end{tabular}\!\!\!\right)\nonumber
\end{equation}

\begin{equation}
\left( \!\!\begin{tabular}{c} 
$\CKP_1$ \cr $\CKP_2$ \cr $\CKJ$ 
\end{tabular} \!\!\! \right)\!\!
=
\!\!
\left(\!\!\begin{tabular}{c}
$\CKCv \dot{x} + \ki\kii \CKSv \CKSx \CKCx\dot{v}$\cr
$\cu^2(x) \dot{v}$\cr
$-\CKSv \dot{x} \!+\! \CKCv \CKSx \CKCx \dot{v}$
\end{tabular}\!\!\right)\!
=
\!\!
\left(\!\!\begin{tabular}{c} 
$\CKCySq \dot{u}$\cr 
$\ki \CKSu \CKSy \CKCy \dot{u} + \CKCu \dot{y}$\cr
$-\CKCu \CKSy \CKCy \dot{u} + \CKSu\dot{y}$
\end{tabular}\!\!\right)
\nonumber
\end{equation}

In the standard Euclidean ${\bf E}^2$, where $\ki\=0, \kii\=1$ and both parallel type coordinates reduce to Cartesian ones,  the CK momenta are in these coordinates: 
\begin{equation}
\left.\CKP_1\right|_{{\bf E}^2}=\dot{x}=\dot{u}, \qquad
\left.\CKP_2\right|_{{\bf E}^2}=\dot{v}=\dot{y}, \qquad
\left.\CKJ\right|_{{\bf E}^2}=x\dot{v}-v\dot{x}=-y\dot{u}+u\dot{y}=x\dot{y}-y\dot{x}\,.
\label{EucP1P2ParCoord}
\end{equation}

\subsection{The curved Kepler problem}

The `curved' Kepler potential in any $\CKspace$ configuration space is:
\begin{equation}
\CKPot_{K} = -\frac{k}{\CKTr}.
\label{CKKepPot}
\end{equation}

In the three Riemannian spaces of constant curvature (CK spaces with $\kii\>0$), this potential follows by enforcing  Gauss law in the corresponding CK three dimensional space, and in particular in the hyperbolic plane it was considered early by Lobachewski himself; see some historical comments in \cite{DoZi91}.  In quantum mechanics on the sphere case this potential was discussed in the paper by Schr\"odinger \cite{Sch40}, shortly to be followed by Infeld and Schild \cite{IS45} who studied the hyperbolic space case. 

Motion in Kepler potential in a {\it curved} configuration space has two trivial constants of motion,  energy and the angular momentum:
\begin{equation}
\begin{tabular}{l}
$E=\frac12 (\CKP_1^2 + \kii \CKP_2^2 + \ki \kii\CKJ^2) - 
\displaystyle\frac{k}{\CKTr} ,\qquad 
\CKJ= \CKSrSq\dot{\phi}$,
\end{tabular}
\label{CKepConsE}
\end{equation}
and there are two additional constants, to be discussed later, which are associated to the separability of the Kepler potential in two  equipara\-bolic coordinate systems (see \cite{CRSS05}). 

In the , the orbit is known to be a conic (for the intrinsic geometry of $\CKspace$), with a focus at the potential origin \cite{CRS07a, CRS07c}. In polar coordinates, with the periastron placed on the half-line $\phi=0$ (hereafter taken as the {\it standard} position of any Kepler orbit), its equation in the CK space $\CKspace$ is:
\begin{equation}
\CKTr=\sqrt{\kii}\CKttt(p)\,\frac{1}{1 + e \CKcc(\phi)} 
\label{CKepOrbit}
\end{equation}
The two geometric parameters determining the conic, i.e., its {\it CK eccentricity} $e$ and {\it semilatus rectum} $p$ are related to the physical conserved quantities {\it energy} $E$ and {\it angular momentum} $\CKJ$ as:
\begin{equation}
\sqrt{\kii}\CKttt(p) = \frac{{\kii} \CKJ^2 }{k}, \qquad e^2 = 1 + \frac{(2E - \ki\kii \CKJ^2)\kii\CKJ^2}{k^2}
\label{CKepRelsEJ_ep}
\end{equation}
Orbits in {\it general} position involve a further constant $\phi_0$: the angular position of the periastron; its equation follows from (\ref{CKepOrbit}) by the trivial replacement $\phi \to \phi-\phi_0$. The constant $\phi_0$ is not essential, as it comes from the central nature of the potential, and will be disregarded from now on. 

We note also a simple relation between $r$ and $u$ along any orbit:
\begin{equation}
\CKTr = \frac{\kii\CKJ^2}{k} - e \CKTu = \sqrt{\kii} \CKttt(p) - e \CKTu 
\label{CKepOrbitRU}
\end{equation}
which follows from the relations (\ref{RelsCoordUV_RPhi}) among coordinates and hold for any CK space; these lead to the $\phi$-dependence of $u, v$ for the orbit in standard position: 
\begin{equation}
\CKTu = \sqrt{\kii} \CKttt(p) \, \frac{\CKCphi}{1 - e \CKCphi}, 
\qquad 
\CKTv = \sqrt{\kii} \CKttt(p) \, \frac{\CKSphi}{1 - e \CKCphi}. 
\end{equation}

\subsection{Levi-Civita regularization of Kepler motion in a curved configuration space}

As the potential is central, the angular momentum constant $\CKJ$ leads to the `curved' version of the {\it law of areas}:
\begin{equation}
\dot{\phi} = \frac{d\phi}{dt} = \frac{\CKJ}{\CKSrSq}, \qquad t = \frac{1}{\CKJ} \int \CKSrSq \,d\phi\,.
\label{CKep_AreasLaw}
\end{equation}

In the euclidean case, time evolution $x(t),\, y(t)$ for cartesian coordinates satisfies a {\it non linear} system of differential equations, which do not allow closed form integration in terms of elementary functions. For elliptic orbits, the problem of motion was solved by Kepler \cite{Ke1609}, essentially  introducing a new parameter along the orbit, the auxiliar angle known as the {\it eccentric anomaly}, instead of taking the polar angle $\phi$, which was known in the classical parlance as the {\it true anomaly}. The eccentric anomaly (which can be also suitably defined for  parabolic and hyperbolic orbits) surprisingly simplifies the situation so as to allow closed elementary expressions. 

The simplification afforded in the Kepler problem by the use of the eccentric anomaly  can be considered as a consequence of a more fundamental structure: when the time $t$ is replaced by another parameter, the {\it (fictitious) Levi-Civita time} $s$ (also known as the Levi-Civita regularization parameter) related to $t$ by the condition $\dot{s}= 1/r$, the equations for the $s$-evolution of $x, y$ become linear \cite{LeCiv06}. If this happens in the Euclidean Kepler problem, a natural question for the `curved' Kepler problem is: does there exists a {\it `curved' regularization parameter} $s$  and some (functions) of coordinates on $\CKspace$ (providing some {\it curved} extension of the cartesian coordinates $x,y$), whose $s$-evolution for the curved Kepler problem is given by a {\it linear system\/}? And to which extent and precisely how other properties already known for the Euclidean Kepler problem still hold for the `curved one'? This is the problem we address in this paper. 

Any {\it ansatz} for the `curved' functions of coordinates and the {\it `curved'} regularization parameter should reduce as $\ki\to0$ to the cartesian coordinates and to the `Euclidean' Levi-Civita parameter determined by $\dot{s}=1/r$. 

Our choice for the curved version of the regularization Levi-Civita parameter is:
\begin{equation}
\dot{s} = \frac{ds}{dt} = \frac{1}{\CKSr\CKCr}, 
\qquad s=\int\frac{1}{\CKSr\CKCr}\,dt\,.
\label{CKep_sDef}
\end{equation}
At first sight, this seems to be an arbitrary choice over other apparently  more natural possibilities, as $\dot{s}=1/\CKTr$ or $\dot{s}=1/\CKSr$; the results will justify however the particular choice made here. 

By combining with the law of areas we get as well:
\begin{equation}
\frac{ds}{d\phi} = \frac{\CKTr}{\CKJ}, 
\qquad s=\frac{1}{\CKJ} \int \CKTr\,d\phi\,.
\label{CKep_EqPhiS}
\end{equation}

Direct computation leads then to the following result: Along any curved Kepler motion, the $s$ evolution of the two coordinate functions $\CKTu, \CKTv$ (\ref{RelsCoordUV_RPhi}) is given by the {\it linear} system
\begin{equation}
\displaystyle\frac{d}{ds} \CKTu = 
            \displaystyle-\frac{k}{\CKJ} \CKTv\,,    \qquad 
\displaystyle\frac{d}{ds} \CKTv = 
            \displaystyle e \CKJ + \frac{k}{\CKJ} \frac{1-e^2}{\kii}\CKTu\,,    
\label{CKep_EqTuTvS}
\end{equation}
hence satisfies the (uncoupled) linear second-order system 
\begin{equation}
\displaystyle\frac{d^2}{ds^2} \CKTu = \displaystyle
               - k e -\frac{k^2}{\CKJ^2} \frac{1-e^2}{\kii}\CKTu \,,  \qquad 
\displaystyle\frac{d^2}{ds^2} \CKTv = 
                \displaystyle  -\frac{k^2}{\CKJ^2} \frac{1-e^2}{\kii}\CKTv \,.  
\end{equation}

The coefficients of $\CKTu, \CKTv$ in the r.h.s.\ are equal in the two equations and the sign of this quantity will determine the nature of the $s$-dependence for the solutions. Expressing this coefficient in terms of the energy and angular momentum through (\ref{CKepRelsEJ_ep}), we obtain:
\begin{equation}
\frac{k^2}{\CKJ^2}\frac{1-e^2}{\kii} = -(2E - \ki\kii \CKJ^2)\,.
\end{equation}
Here we see that in the curved case, the quantity governing the character  of the evolution is not $-2E$ as it was in the Euclidean case \cite{MilnorGKP} but rather the combination $-(2E - \ki\kii \CKJ^2)$, which has, further to the energy, a contribution from the angular momentum. If we define $\sLab$ as
\begin{equation}
\sLab:= - (2E - \ki\kii \CKJ^2)\,, 
\label{CKepDefslab}
\end{equation}
and arrange the constant terms, the second order system is:
\begin{equation}
\displaystyle\frac{d^2}{ds^2} \Big(\CKTu + \frac{ke}{\sLab}\Big)  = 
                 \displaystyle-\sLab \Big(\CKTu + \frac{ke}{\sLab}\Big)\,,
\qquad
\displaystyle\frac{d^2}{ds^2} \CKTv  = -\sLab \CKTv \,, 
\end{equation}
whose {\it general} solution, for either $\Big(\CKTu + \frac{ke}{\sLab}\Big)$ or $\CKTv$ as functions of $s$ is, after (\ref{DifEqDefSinCos}), a linear combination $A \CKcLab(s) + B \CKsLab(s)$ of the two basic solutions $\CKcLab(s), \CKsLab(s)$. This is the more general solution, even when $\sLab=0$, for then  it reduces to $A + B s$. The symbol $\sLab$ has been chosen to underline its natural link with $s$, as the `CK label' of the Levi-Civita parameter $s$ (just as the lenghts $u, r$ have `label' $\ki$ and the angles $\phi$ have `label' $\kii$, in the sense any appearance of these quantities is through the corresponding `labelled' functions). 

Now for the evolution along the orbit in the standard position (with $\phi_0=0$), at the periastron $v$ must vanish and $u$ must be extremal; if the origin of $s$ is chosen also at the periastron, enforcing the previous  conditions when $s=0$, the expressions for $\CKTu, \CKTv$ are narrowed down to:
\begin{equation}
\CKTu = A \CKcLab(s) - \frac{ke}{\sLab}, \qquad \CKTv = B \CKsLab(s)
\end{equation}
and finally $A, B$ can be related to the physical constants along the Keper motion by noticing that at the periastron, where $s=0$ one must have $u=r_{per}$, and hence by equating 
$\left.\CKTu\right|_{s=0}=A-{ke}/{\sLab}$ to $\CKt(r_{per})=\sqrt{\kii} \CKttt(p)/({1+e)}$ taken from (\ref{CKepOrbit}) and using (\ref{CKepDefslab}) we get after some algebra $A={k}/{\sLab}$. Hence:
\begin{equation}
\CKTu ={k} \,\frac{\CKcLab(s) - e}{\sLab}\,.
\label{CKepTu_s}
\end{equation}

To determine the constant $B$, let $s_{sl}$ be the value of the parameter $s$ at the semilatus rectum of the orbit. At this point $u=0$ and $v=p$. We have then two equations:
\begin{equation}
{k} \,\frac{\CKcLab(s_{sl}) - e}{\sLab}=0, \qquad B \CKsLab(s_{sl}) = \CKttt(p)
\end{equation}
and elimination of $s_{sl}$ through the basic identity 
$\CKC^2_\sLab(s) + \sLab \CKS^2_\sLab(s) = 1$ 
leads, after some work to $B =\pm \CKJ$. The choice of sign corresponds to the sense of motion along the orbit and in what follows we will choose $B=\CKJ$. Thus: 
\begin{equation}
\CKTv = \CKJ \CKsLab(s) 
\label{CKepTv_s}\end{equation}

Up to now we have found the $s$ dependence of $u$ and $v$ (through $\CKTu, \CKTv$). The dependence of the radial coordinate $r$ on $s$ follows directly from (\ref{CKepOrbitRU}):
\begin{equation}
\CKTr = \frac{\kii \CKJ^2}{k} - {k e} \,\frac{\CKcLab(s) - e}{\sLab} =
        k \, \frac{1 - e  \CKcLab(s)}{\sLab}
\label{CKepTr_s}\end{equation}
where we used the relations (\ref{CKepRelsEJ_ep}) which are better rewritten in terms of $\sLab$ as
\begin{equation}
\sqrt{\kii}\CKttt(p)=k \frac{1-e^2}{\sLab}\,.\qquad
(1-e^2) = \frac{\kii\sLab\CKJ^2}{k^2}\,.
\label{CKepRelEP2}
\end{equation}
In both equations, when $\sLab\to 0$ or when $\kii\to 0$, then $e\to 1$ while  the quotient $(1-e^2)/(\sLab\kii)=\CKJ^2/k^2$ remains well defined and depend only on the angular momentum Notice that in spaces with a degenerate metric ($\kii=0$), all Kepler orbits have $e=1$, in this case the orbit equation  (\ref{CKepOrbit}) should be rewitten in terms of the versed sine of the angle $\phi$. 

The use the Levi-Civita parameter enables the finding of closed exact and smooth expressions for the  $s$-dependence of $u, v$ and $r$ in the general `curved' case almost as easily in the Euclidean case. 
For the relation among $s$ and  the angular coordinate $\phi$ we may use (\ref{CKep_EqPhiS}) and replace $\CKTr$ taken either from (\ref{CKepOrbit}) or from (\ref{CKepTr_s}) to find two alternative forms:
\begin{equation}
\frac{ds}{d\phi} = \frac{k}{\CKJ}\,\frac{1 - e  \CKcLab(s)}{\sLab} = 
\frac{\CKJ}{k}\,\frac{\kii}{1+ e \CKCphi}
\label{CKepS_phi}
\end{equation}
Each of these forms allows closed integration by using the formula: 
\begin{equation}
\int\frac{1}{1- e \CKC_ \sLab(s)} \, ds = -\frac{2}{\sqrt{ e ^2-1}} \CKArcT_{-\sLab}\left( {\frac{ e +1}{\sqrt{ e ^2-1}}}\CKT_ \sLab\left(\frac{s}{2}\right)\right)
\label{PrimExtLambda} 
\end{equation}
Integrating (\ref{CKepS_phi}) (notice that the integration can be done in two ways as the equation allows two separated forms) and using (\ref{CKepRelEP2}) gives: 
\begin{equation}
\sqrt{\frac{\kii}{-\sLab}}\frac{\phi}{2} = 
\CKArcT_{-\sLab}\left( {\frac{ e +1}{\sqrt{ e ^2-1}}}\CKT_ \sLab\left(\frac{s}{2}\right)\right)
\label{CKepTphiH_s1}
\end{equation}
leading after some manipulations to the $s$-dependence of the angular coordinate $\phi$: 
\begin{equation}
\sqrt{\kii}\CKtt\left(\frac{\phi}{2}\right) =
 (1+e)\sqrt{\frac{\sLab}{1-e^2}} \CKtLab\left(\frac{s}{2}\right)
\label{CKepTphiH_s}
\end{equation}
which remains meaningful even when $\sLab=0$, as clear from (\ref{CKepRelEP2}). 
This formula does not involve explicitly the curvature $\ki$ of the configuration space. For the standard Euclidean case, where $\kii=1$, the l.h.s.\  will be $\tan(\phi/2)$, so this formula ressembles closely the well known Euclidean relation between the angle $\phi$ and the eccentric anomaly; indeed there is a direct relation, in the Euclidean case,  between  $s$ and the eccentric anomaly, as we will see  in the last section. 

The equation (\ref{CKepS_phi}) displays a curious symmetry among $\CKCphi$ and $\CKcLab(s)$, which can be enhanced by writing it in the form:
\begin{equation}
\frac{1 - e  \CKcLab(s)}{\sLab} \, \frac{1+ e \CKCphi}{\kii} = \frac{\CKJ^2}{k^2} =
\frac{1-e^2}{\sLab \kii}
\label{CKepRelSigmaPhi}
\end{equation}

Closed formulas can also be obtained for the three basic (cosine, sine and tangent) trigonometric functions of the angular coordinate; the most direct way is to recall the relations (\ref{RelsCoordUV_RPhi}) which hold for the coordinates in any CK  space obtaining:
\begin{equation}
\begin{array}{ll}
\CKCphi = & \displaystyle\frac{\CKcLab(s) - e}{1 - e  \CKcLab(s)} 
\\[8pt]
\CKSphi = & 
    \displaystyle\frac{\CKJ}{k} \frac{\sLab \CKsLab(s)}{1 - e  \CKcLab(s)} =
    \sqrt{\frac{1-e^2}{\kii\sLab}}\frac{\sLab\CKsLab(s)}{1 - e  \CKcLab(s)}\\[8pt] 
\CKTphi = & 
    \displaystyle\frac{\CKJ}{k} \frac{\sLab \CKsLab(s)}{\CKcLab(s) - e} =
    \sqrt{\frac{1-e^2}{\kii\sLab}}\,\frac{\sLab\CKsLab(s)}{\CKcLab(s) - e}
    \end{array}
	\label{CKepPhi_s}
\end{equation}
(of course, starting from these expressions, the relation (\ref{CKepTphiH_s}) can be also derived by direct computation instead of by integration). 

We already have obtained the $s$-dependence of the radial coordinate. Let us ask now for the time $t$ as a function of $s$. By starting from the definition (\ref{CKep_sDef}), using the identity $1/\CKCrSq=1+\ki\CKTrSq$ and the $s$-dependence of $\CKTr$ we obtain: 
\begin{equation}
\frac{ds}{dt} = \frac{1}{\CKSr \CKCr} = \frac{1}{\CKTr \CKCrSq} = 
\frac{1+ \ki \CKTrSq}{\CKTr} = \frac{1 + 
\ki \Big({k \,\frac{1 - e  \CKcLab(s)}{\sLab}} \Big)^2}
{k \, \frac{1 - e  \CKcLab(s)}{\sLab}}
\end{equation}
so that, with the conventional choice for the time origin $t=0$ when the particle is at the periastron: 
\begin{equation}
t = k \int \frac{\frac{1}{\sLab} \big(1 - e  \CKcLab(s)\big)}{1 + \ki \frac{k^2}{\sLab^2} \big(1 - e  \CKcLab(s)\big)^2} \, ds
\label{CKIntT}
\end{equation}
The integration can be done by decomposing in simple fractions, and leads to:
\begin{equation}
t(s) = \frac{1}{2 e \sqrt{-\ki}}\int \left\{ 
\frac{\alpha - e}{ 1 - \alpha \CKcLab(s)} + \frac{\beta + e}{ 1 + \beta \CKcLab(s)}
\right\} \, ds
\end{equation}
where $\alpha, \beta$ are dimensionless quantities:
\begin{equation}
\alpha=\frac{k e \sqrt{-\ki}}{\sLab + k \sqrt{-\ki}}, \qquad 
\beta=\frac{k e \sqrt{-\ki}}{\sLab - k \sqrt{-\ki}}
\label{CKepDefAlphaBeta}
\end{equation}
which by using (\ref{PrimExtLambda}) can be expressed in completely closed form: 
\begin{equation}
t(s) = \frac{1}{e \sqrt{-\ki}} \left\{
\frac{\alpha-e}{\sqrt{\alpha^2-1}} \CKArcT_{-\sLab}\left( 
{\frac{\alpha+1}{\sqrt{\alpha^2-1}}}\CKT_\sLab\left(\frac{s}{2}\right)\right)
+
\frac{\beta+e}{\sqrt{\beta^2-1}} \CKArcT_{-\sLab}\left( 
{\frac{\beta-1}{\sqrt{\beta^2-1}}}\CKT_\sLab\left(\frac{s}{2}\right)\right)
 \right\} 
\end{equation}
By noticing $\alpha-e=-\frac{\sLab}{k\sqrt{-\ki}}\alpha$, $\beta+e=\frac{\sLab}{k\sqrt{-\ki}}\beta$, the previous expression can be also given as (notice the dissapearance of $e$ in the previous $\alpha-e$, $\beta+e$ terms, the relative minus sign and the change in the global prefactor in the new expression):
\begin{equation}
t(s) = \frac{\sLab}{e k \ki} \left\{
\frac{\alpha}{\sqrt{\alpha^2-1}} \CKArcT_{-\sLab}\left( 
{\frac{\alpha+1}{\sqrt{\alpha^2-1}}}\CKT_\sLab\left(\frac{s}{2}\right)\right)
-
\frac{\beta}{\sqrt{\beta^2-1}} \CKArcT_{-\sLab}\left( 
{\frac{\beta-1}{\sqrt{\beta^2-1}}}\CKT_\sLab\left(\frac{s}{2}\right)\right)
 \right\} 
 \label{CKepT_s}
\end{equation}

Summing up: closed expressions for the coordinates $u, v, r, \phi$ and the time $t$ along a Kepler orbit in a curved configuration space have been obtained. Together with (\ref{CKepTu_s}, \ref{CKepTv_s}, \ref{CKepTr_s}), this expression completely solves the configuration space with any constant curvature and any signature, and for any value of the constant $\sLab$. 

\section{The Kepler motion in a curved configuration space as a geodesic flow}

Now we want to explore whether or not  the well known relation among the Kepler motions with energy $E$ in Euclidean space and  a geodesic flow in a space of constant curvature $-2E$ (and definite positive metric) extends for the `curved' Kepler motion in any $\CKspace$. A constructive way to establish this connection is to look to the Kepler evolution not in the configuration space $\CKspace$ but in an auxiliary space, the space of Cayley-Klein momenta $\CKP_1, \CKP_2$. 

The `curved' Kepler problem has, further to energy and angular momentum, two additional constants of motion, which can be considered as the components of a single vector, the {\it eccentricity} or Hamilton vector, which is related to the Laplace-Runge-Lenz vector (see \cite{CRS07c} for comments on the relation between the eccentricity vector and the Laplace-Runge-Lenz vector in the `curved' case and \cite{Go75PRLV, Go76PRLV, LeFl03} for additional information in the Euclidean case). In any configuration space $\CKspace$, \cite{CRS07a, CRS07c} these constants are:
\begin{equation}
\CKE_{01} = \CKJ \CKP_1 + k \CKSphi, \qquad 
\CKE_{02} = \CKJ \CKP_2 + k \CKVphi
\label{CKepEccentVect}
\end{equation}
The existence of these Kepler first integrals can be seen as a consequence of the separability of the curved Kepler potential in two systems of `parabolic' coordinates in the curved configuration space (with a focus at the potential origin) \cite{CRSS05}. The particular form of the constants (\ref{CKepEccentVect}) has a geometric consequence: the `momentum hodograph' is a `cycle' in the plane of momenta $\CKP_1, \CKP_2$, a plane whose metric has signature type $\kii$ (this is suggested in (\ref{CKepConsE})). 
Taking the constancy of (\ref{CKepEccentVect}) as the departure point would be a modern, natural choice. For the Euclidean Kepler problem, the circle character of hodographs was first stated explicitly by Hamilton, who actually derived things the other way round: from a suitable rewriting of Newton's equations, the circle character of the (velocity) hodograph follows, and this leads to the two new constants of motion specific to the Kepler problem, the Hamilton eccentricity vector. So let us mimic, for the general CK space, the path trodden by Hamilton in the Euclidean Kepler problem \cite{HamiltonHodog} and let us try to {\it derive} the constancy of (\ref{CKepEccentVect}), keeping in mind an important fact: Hamilton considered the {\it velocity} hodograph vector, and we are considering the `momentum hodograph' $\CKP_1, \CKP_2$; this will be of some relevance later.

When a particle moves in $\CKspace$ following Kepler evolution, angular momentum $\CKJ$ is constant, but $\CKP_1, \CKP_2$ do depend on time. Their time evolution 
is the extension to the space $\CKspace$ of the original form of Newton's equations: the rate of change of momentum is equal to the force. This `force' is a vector under rotations around the origin (hence, a vector in a flat plane with signature type $\kii$), and its components are $(\CKF_1, \CKF_2) = -k(\CKCphi, \CKSphi)/\CKSrSq$, whose radial dependence $1/\CKSrSq$  follows directly from the Gauss law in a 3d space of constant curvature $\ki$ (where the area of the sphere of radius $r$ grows as $\CKSrSq$; notice also $d(k/\CKTr)/dr=-k/\CKSrSq)$). Hence we may write: 
\begin{equation}
\dot{\CKP_1} = \frac{d\CKP_1}{dt} = -\frac{k}{\CKSrSq} \CKCphi \,, 
\qquad
\dot{\CKP_2} = \frac{d\CKP_2}{dt} = -\frac{k}{\CKSrSq} \CKSphi \,. 
\label{CKepDMomentumTime}
\end{equation}
Elimination of $t$ using of the law of areas leads to: 
\begin{equation}
\frac{d \CKP_1}{d\phi}= -\frac{k}{\CKJ}\CKCphi, 
\qquad
\frac{d \CKP_2}{d\phi}= -\frac{k}{\CKJ}\CKSphi, 
\label{CKepDMomentumPhi}
\end{equation}
a system whose integration is trivial,  leading precisely to (\ref{CKepEccentVect}). Had we started from the constancy of the two components of the eccentricity vector, the equations (\ref{CKepDMomentumTime})  and (\ref{CKepDMomentumPhi}) would follow simply by differentiating  (\ref{CKepEccentVect}) with respect to time and using the law of areas (\ref{CKep_AreasLaw}). The orbit follows directly by enforcing the relation between the three CK momenta which generalizes the euclidean $\CKJ=x\CKP_2 - y \CKP_1$ \cite{CRS07a, CRS07c}. 

This means that the `momentum hodograph' curve is always a `cycle' in the $\CKP_1, \CKP_2$ space relatively to the (flat) metric with signature type $\kii$, based in the quadratic form $d\wp^2:=d\CKP_1^2 + \kii \CKP_2^2$. (We recall cycles are defined as the curves with constant geodesic curvature; in the Euclidean plane these are circles, with straight lines as limiting cases). Of course in the Euclidean Kepler problem, the two momenta $\CKP_1, \CKP_2$ are simply (\ref{EucP1P2ParCoord}) equal to the two cartesian components of the ordinary velocity vector. Thus the Hamilton result for the Euclidean Kepler problem, formulated in terms of {\it velocity hodographs}, could equivalently be restated in terms of {\it momentum hodographs} without any change whatsoever. But this equality does not extend for curved spaces. 
No simple extension of the Hamilton result should be therefore expected for the `velocity' vector in the `curved' Kepler case because the velocity is a tangent vector to each point of the trajectory, which would require some kind of transport to a common origin to make sense of a `curved velocity hodograph'; this would mean some further arbitrary choice as to the path to follow for performing the transport. On the other hand, the `momentum hodograph' makes sense in any curved configuration space, without any further assumption, because the Cayley-Klein momentum $\CKP_1, \CKP_2$, already lives in a Lie algebra and can be considered as a vector attached to the origin of a linear space without any need to perform any kind of `transport'.

Hence we may state as the first main result

\medskip
{\bf Theorem} [`curved' version of Hamilton's] {\sl As $t$ varies, the CK momentum vector $\CKP_1, \CKP_2$ of a particle undergoing Kepler evolution in a space $\CKspace$, moves along a `cycle' in the `momentum plane' whose metric $d\wp^2 = d\CKP_1^2 + \kii d\CKP_2^2$ is flat and of signature type $\kii$}. 

\medskip

Now we ask about the relation among the Levi-Civita parameter $s$ with the metric 
$d\wp^2$ in this result. The (square of the) `norm' of $\dot{\pmb{\CKP}} = {d\pmb{\CKP}}/{dt} \equiv (d\CKP_1/dt, d\CKP_2/dt)$ seen as a {\it vector in a 2d plane} whose (flat) metric has signature $\kii$ is:
\begin{equation}
\dot{\pmb{\CKP}} \cdot \dot{\pmb{\CKP}} = (\dot{\CKP_1})^2 + \kii (\dot{\CKP_2})^2 = 
\frac{k^2}{\CKS^4_{\ki}(r)}\,.
\end{equation}
Consider now the $s$-dependence of momenta, related to the time dependence as:
\begin{equation}
\frac{d\pmb{\CKP}}{ds} = \frac{d\pmb{\CKP}}{dt} \,\frac{dt}{ds} = \CKSr\CKCr \,\frac{d\pmb{\CKP}}{dt}\,,
\end{equation}
so that for the `norm' of the vector ${d\pmb{\CKP}}/{ds}$ we have:
\begin{equation}
\frac{d\pmb{\CKP}}{ds} \cdot \frac{d\pmb{\CKP}}{ds} =
\frac{(d\CKP_1)^2 + \kii (d\CKP_2)^2}{ds^2} = 
\CKSrSq\CKCrSq \frac{k^2}{\CKS^4_{\ki}(r)} = 
\frac{k^2}{\CKTrSq}\,.
\end{equation}
Conservation of energy (\ref{CKepConsE}) allows to draw $k/\CKTr$ in terms of momenta, \begin{equation}
\frac{k}{\CKTr} = E - \smallonehalf \ki\kii \CKJ^2 - \smallonehalf\left\{ (\CKP_1)^2 + \kii (\CKP_2)^2 \right\}=
- \smallonehalf\left\{ (\CKP_1)^2 + \kii (\CKP_2)^2 + \sLab \right\}
\end{equation}
where again the quantity $\sLab$ appears. By combining these relations, we get: 
\begin{equation}
ds^2 = \frac{4 \left\{ (d\CKP_1)^2 + \kii (d\CKP_2)^2 \right\}}{\big( (\CKP_1)^2 + \kii (\CKP_2)^2  + \sigma \big)^2} =
\frac{4 }{\big( (\CKP_1)^2 + \kii (\CKP_2)^2  + \sigma \big)^2} d\wp^2 
\label{CKepLCMetricP}
\end{equation}
Hence, if (\ref{CKepLCMetricP}) is considered as the {\it definition} of a metric  in the momentum plane, the increase of the Levi-Civita parameter along any Kepler orbit is equal to the `lenght' provided by this `Levi-Civita metric' (\ref{CKepLCMetricP}) along the curve described by the evolution of the CK momenta $\CKP_1, \CKP_2$. 
The metric $ds^2$ on the space of all CK momenta admissible for a Kepler motion with fixed $\sLab$, does depend on $\sLab$. It is {\it different} from the flat momentum plane metric $d\wp^2$ implicit in the previous theorem,  but both metrics are conformal, with a conformal factor depending on $\sLab$. By direct computation it follows that the curvature of this (Levi-Civita) metric is {\it constant} and equal to $\sLab$. The checking for the last property could be even bypassed by suitably adapting and extending (so as to include also the parameter $\kii$) the idea used by Milnor for the Kepler problem in Euclidean space: consider the `inverted' momenta vector $\pmb{\CKW}$, which may be called the  `slowmentum' and is defined as: 
\begin{equation}
\CKW_1 = \frac{\CKP_1}{\CKP_1^2 + \kii \CKP_2^2}, \qquad 
\CKW_2 = \frac{\CKP_2}{\CKP_1^2 + \kii \CKP_2^2}
\label{CKepDefW}
\end{equation}
Straightforward computations lead to several simple relations, whose standard $\kii=1$ versions are well known:
\begin{equation}
\CKW_1^2 + \kii \CKW_2^2  = \frac{1}{\CKP_1^2 + \kii \CKP_2^2}, \qquad
d\CKW_1^2 + \kii d\CKW_2^2 = \frac{d\CKP_1^2 + \kii d\CKP_2^2}{\big( \CKP_1^2 + \kii \CKP_2^2 \big)^2}
\label{CKepRelV_W}
\end{equation}
allowing to obtain the expression of the Levi-Civita metric in terms of the `slowmentum':
\begin{equation}
ds^2 = \frac{ 4 \left\{ (d\CKW_1)^2 + \kii (d\CKW_2)^2 \right\}}{\big( 1 + \sLab \big( \CKW_1^2 + \kii \CKW_2^2\big) \big)^2}\,,
\end{equation}
which is precisely the form for the metric in a CK space with constant curvature $\sLab$ and signature type determined by $\kii$ in Riemann's normal coordinates. 

At this point, we expect that the well known connection between the Kepler motion in Euclidean space and the geodesic flow on constant curvature spaces extend so as to include also the Kepler problem on a `curved' configuration space. This connection can be established following two paths which are however very closely related: either disclosing a stereographic projection implicit in the expressions we have found, or by a direct analysis of the geometry of cycles in the $\CKP_1, \CKP_2$-plane (compare Milnor \cite{MilnorGKP} and Anosov \cite{AnosovGKP} in the Euclidean case). In the next section we complete the details. 

\subsection{Kepler motion as stereographic projection of free motion}

We start from the closed expressions for the `momentum hodograph' corresponding to the CK orbit in the standard position. These follow by integration of the system (\ref{CKepDMomentumPhi}) and with the correct choice of the integration constants they are:
\begin{equation}
\CKP_1(\phi) = -\frac{k}{\CKJ} \CKSphi, \qquad 
\CKP_2(\phi) = \frac{k}{\kii\CKJ} \Big( e + \CKCphi \Big)\,.
\label{CKepP1P2_Phi}
\end{equation}
This curve is a `cycle' in the $(\CKP_1, \CKP_2)$-plane, whose `radius' is $k/(\sqrt{\kii}\CKJ)$ and whose center lies on the $\CKP_2$ axis, at the point $(\CKP_1, \CKP_2)=(0, ke/(\kii\CKJ))$. This also determines the values of the constants $\CKE_{01}, \CKE_{02}$ for a Kepler orbit in standard position:
\begin{equation}
\CKE_{01}=0, \qquad \CKE_{02}= \frac{k(1+e)}{\kii}\,.
\end{equation}
The $s$-dependence of the CK momenta $\CKP_1, \CKP_2$ can be found by replacing (\ref{CKepPhi_s}) in (\ref{CKepP1P2_Phi}): 
\begin{equation}
\CKP_1(s) = - \frac{\sLab\CKsLab(s)}{1 - e  \CKcLab(s)}, 
\qquad
\CKP_2(s) = \frac{\CKJ}{k}\, \frac{\sLab \CKcLab(s)}{1 - e  \CKcLab(s)}\,,
\end{equation}
where by direct computation:
\begin{equation}
\CKP_1^2(s) + \kii \CKP_2^2(s) =  \sLab\, \frac{1 + e  \CKcLab(s)}{1 - e  \CKcLab(s)}\,.
\end{equation}

For the `inverted' momentum or slowmentum vector, the $s$-evolution is: 
\begin{equation}
\CKW_1(s) = -\, \frac{\CKsLab(s)}{1 + e  \CKcLab(s)}, 
\qquad
\CKW_2(s) = \frac{\CKJ}{k}\, \frac{\CKcLab(s)}{1 + e  \CKcLab(s)}
\label{CKepInverMomS}
\end{equation}
and either by direct computation or using (\ref{CKepRelV_W}):
\begin{equation}
\CKW_1^2(s) + \kii \CKW_2^2(s) =  \frac{1}{\sLab}\, \frac{1 - e  \CKcLab(s)}{1 + e  \CKcLab(s)}
\end{equation}
The equations (\ref{CKepInverMomS}) can be directly recognized as the stereographic projection of a free motion in a space $\CKspaceKepMom$ with constant curvature $\sLab$ and metric of signature type $\kii$ onto a space with a flat metric of signature type $\kii$. Checking this statement involves a  extension of the standard stereographic projection (where $\kii=1$) to a $\kii$-general situation, making thus sense for all CK spaces. 

The classical stereographic projection maps the standard sphere ${\bf S}^2$ of constant curvature $1$ (realized as the submanifold $(s^0)^2 + (s^1)^2 + (s^2)^2 = 1$ in an auxiliar ambient space $(s^0, s^1, s^2)$) on the flat plane $s^0=1, s^1\equiv w^1,  s^2\equiv w^2$ living in this auxiliar space, by projecting the sphere from the `South pole' $(s^0, s^1, s^2)=(-1, 0,0)$; this mapping is well known to preserve angles. 
A geometrically similar construction maps {\it any CK space} $\CKspace$, realized as the `sphere' $(s^0)^2 + \ki(s^1)^2 + \ki\kii(s^2)^2 = 1$ in an auxiliar ambient space $(s^0, s^1, s^2)$, over the flat plane $s^0=1, s^1\equiv w^1,  s^2\equiv w^2$, by projecting the `sphere' over the plane from the `South pole' $(s^0, s^1, s^2)=(-1, 0,0)$. This general stereographic projection is described, in terms of the ambient coordinates in ${\bf S}^2$ by the map:
\begin{equation}
\left( \!\!\begin{tabular}{c} 
$s^0$ \cr $s^1$ \cr $s^2$ 
\end{tabular} \!\!\! \right) \to 
\left( \!\!\begin{tabular}{c} 
$w^1=\displaystyle\frac{s^1}{1+s^0}$ \cr 
$w^2= \displaystyle\frac{s^2}{1+s^0}$
\end{tabular} \!\!\! \right)\,,
\label{CKStereoPro}
\end{equation}
an expression which do not depends explicitly on $\ki, \kii$. In this form, the stereographic projection makes sense for all CK spaces, either with  Riemannian or Lorentzian signature type. For ${\bf H}^2$, realized in the Weierstrass ambient space model, this stereographic projection provides the Poincar\'e conformal disc model of the hyperbolic plane. 

Consider now in the CK space ${\bf S}^2_{\sLab[\kii]}$  with curvature $\sLab$ and metric of signature type $\kii$, the `fiducial' geodesic described in ambient space coordinates as:
\begin{equation}
(s^0, s^1, s^2) = (\CKcLab(s),-\CKsLab(s),0)
\end{equation}
which is the `basic' line $l_1$ along the direction $1$ traversed negatively starting from the origin point $(1,0,0)$. Then move this geodesic to a new position $l_1'$ by a translation along the orthogonal line $l_2$ with an amount $\epsilon$; we recall that $\epsilon$ has a label $\sLab\kii$, just as the analogous quantity $v$ has label $\ki\kii$ in $\CKspace$. This translation map the fiducial geodesic $l_1$ into some other member $l_1'$ of the set of geodesics orthogonal to the basic line $l_2$. This one-dimensional family of geodesics is described in the ambient space by letting a general translation along the $l_2$ line to act on the fiducial geodesic by matrix multiplication:
\begin{equation}
\left( \!\!\begin{tabular}{c} 
$\CKcLab(s)$ \cr $-\CKsLab(s)$ \cr $0$ 
\end{tabular} \!\!\! \right) \to 
\left(\begin{array}{ccc}
\CKC_{\sLab\kii}(\epsilon)&0&-\sLab\kii\CKS_{\sLab\kii}(\epsilon) \cr 
0&1&0\cr 
\CKS_{\sLab\kii}(\epsilon)&0&\CKC_{\sLab\kii}(\epsilon) \end{array}\right)
\left( \!\!\begin{tabular}{c} 
$\CKcLab(s)$ \cr $-\CKsLab(s)$ \cr $0$ 
\end{tabular} \!\!\! \right) 
=
\left( \!\!\begin{tabular}{c} 
$\CKC_{\sLab\kii}(\epsilon)\CKcLab(s)$ \cr $-\CKsLab(s)$ \cr $\CKS_{\sLab\kii}(\epsilon)\CKcLab(s)$ 
\end{tabular} \!\!\! \right) 
\end{equation}
and this geodesic $l_1'$ in $\CKspaceKepMom$ is mapped, upon stereographic projection (\ref{CKStereoPro}) into the curve:
\begin{equation}
\CKW ^1 = -\frac{\CKsLab(s)}{1+\CKC_{\sLab\kii}(\epsilon)\CKcLab(s)}, \qquad
\CKW ^2 = \frac{\CKS_{\sLab\kii}(\epsilon)\CKcLab(s)}{1+\CKC_{\sLab\kii}(\epsilon)\CKcLab(s)} 
\end{equation}
to be compared with (\ref{CKepInverMomS}); these coincide provided the two identification conditions: 
\begin{equation}
\CKC_{\sLab\kii}(\epsilon) \leftrightarrow e, \qquad 
\CKS_{\sLab\kii}(\epsilon) \leftrightarrow \frac{\CKJ}{k} = \sqrt{\frac{\sqrt{\kii}\CKttt(p)}{{\kii}k}}
\label{CKepIdentEpsilonEP}
\end{equation}
are consistent. This requires
\begin{equation}
\CKC^2_{\sLab\kii}(\epsilon) + \sLab\kii\CKS^2_{\sLab\kii}(\epsilon) = 1
\end{equation}
and by direct checking we find this is satisfied, because this is simply (\ref{CKepRelsEJ_ep}) in disguise. 
The position of any geodesic in ${\bf S}^2_{\sLab[\kii]}$ can be described by two parameters: an orientation angle $\phi_0$ and an `impact parameter' relative to the origin. The angle turns out to be precisely equal to the angle $\phi_0$ determining the orientation of the general Kepler conic in the configuration space. The `impact parameter' $\epsilon$ will contain, simultaneously, information on the eccentricity of the Kepler orbit and on the angular momentum (for the fixed value of the constant $\sLab$). In particular, geodesics with impact parameter equal to zero correspond to collision orbits with $\CKJ=0$; from the geometric picture, as 
the symmetry group of ${\bf S}^2_{\sLab[\kii]}$ acts transitively on points and on either type (`time' or `space' like when $\kii<0$) geodesics, it follows that the collision orbits receive, alike any other orbit, a regular description within this scheme.

The picture is the following: On a CK configuration space $\CKspace$ with constant curvature $\ki$ and metric of signature type $\kii$, the subset of all Kepler motions for which the quantity $\sLab=-(2E-\ki\kii\CKJ^2$) has a fixed value can be realized as the geodesic flow on a CK space ${\bf S}^2_{\sLab[\kii]}$ of constant curvature $\sLab$ and metric of the signature type $\kii$. When a (fictitious) point on this space moves with unit speed along a geodesic $l_1'$ obtained from the `basic' geodesic $l_1$ by a translation of amount $\epsilon$ along $l_2$, its stereographic projection on the `inverted' momentum space corresponds to a (slowmentum hodograph) of the Kepler motion along a orbit in the  standard position, with eccentricity, semilatus rectum and angular momentum determined from $\epsilon$ by (\ref{CKepIdentEpsilonEP}). The arc lenght along the geodesic $l_1'$ is  the `curved' Levi-Civita parameter.

Let us rewrite the results of the previous section, in terms of the identification (\ref{CKepIdentEpsilonEP}):
\begin{equation}
\Big(\CKTu\Big) (s) = k \frac{\CKC_{\sLab}(s)-\CKC_{\sLab\kii}(\epsilon)}{\sLab}, 
\qquad
\Big(\CKTv\Big) (s) = k \CKS_{\sLab\kii}(\epsilon)\CKS_{\sLab}(s), 
\label{CKep_UV_sEpsilon} 
\end{equation}
\begin{equation}
\Big(\CKTr\Big) (s)  = k \frac{1- \CKC_{\sLab\kii}(\epsilon)\CKC_{\sLab}(s)}{\sLab}, 
\qquad
\Big(\CKTphi\Big) (s)  = k \CKS_{\sLab\kii}(\epsilon)\CKS_{\sLab}(s) 
\frac{\sLab}{\CKC_{\sLab}(s)-\CKC_{\sLab\kii}(\epsilon)}
\label{CKep_RPhi_sEpsilon} 
\end{equation}

The structure of these expressions is neat, and their genericity must be emphasized: they hold for any Kepler motion, with any $\sLab$, in any $\CKspace$. 
All involve as a factor the strenght of the Kepler coupling constant $k$. In all cases there are two further {\it independent} variables, which can be identified to parallel `1' coordinates in ${\bf S}^2_{\sLab[\kii]}$. One of these coordinates corresponds to the choice of a standard geodesic (the `impact parameter' $\epsilon$) and the other is the arc lenght parameter along the geodesic. The presence of explicit $\sLab$ in the denominators would seem to imply some difficulties when $\sLab \to 0$, but this is not so. As a consequence of the behaviour of the Cosine and Sine functions, the vanishing of $\sLab$ is automatically accompanied by the vanishing of $1-\CKC_{\sLab}(s)$. 
By introducing the {\it versed sine} $\CKV_{\sLab}(s)$, defined for nonvanishing $\sLab$ as: 
\begin{equation}
\CKV_{\sLab}(s) = \frac{1-\CKC_{\sLab}(s)}{\sLab}, 
\end{equation}
then when $\sLab\to0$ this function reduce to $s^2/2$ and should be considered as the next natural stages in the `curved' analogues of $s^2/2$ just as $\CKC_{\sLab}(s)$ and $\CKS_{\sLab}(s)$ are the `curved' analogues of the functions $1$ and $s$. When (\ref{CKep_UV_sEpsilon}, \ref{CKep_RPhi_sEpsilon}) are reexpressed in terms of the functions $\CKV_{\sLab}(s)$ and $\CKV_{\sLab\kii}(\epsilon)$ we obtain a description of the $s$--evolution
which covers the Kepler motion on {\it any} CK space (with any constant curvature and any signature type)  and which makes furthermore sense for {\it any} value of the combination $\sLab$ of  energy and angular momentum: 
\begin{equation}
\Big(\CKTu\Big)(s) = k \Big( \kii\!\CKV_{\sLab\kii}(\epsilon)-\CKV_{\sLab}(s) \Big) \,,
\qquad
\Big(\CKTv\Big)(s) = k \CKS_{\sLab\kii}(\epsilon)\CKS_{\sLab}(s) \,,
\end{equation}
\begin{equation}
\Big(\CKTr\Big)(s) = k \Big( \kii\!\CKV_{\sLab\kii}(\epsilon) + \CKC_{\sLab\kii}(\epsilon)\CKV_{\sLab}(s) \Big) \,, 
\qquad
\Big(\CKTphi\Big) (s) = k \frac{\CKS_{\sLab\kii}(\epsilon)\CKS_{\sLab}(s)}
{\kii\!\CKV_{\sLab\kii}(\epsilon)-\CKV_{\sLab}(s)}
\end{equation}

In particular, orbits with $\sLab=0$, which are in some aspects analogues in the `curved' configuration space of the (non-generic) parabolic orbits in Euclidean space, are described as the particular case $\sLab=0$ of the previous generic expressions, which simplify in this case to:
\begin{equation}
\Big(\CKTu\Big)(s) = k \left( \frac{\kii\epsilon^2}{2}-\frac{s^2}{2} \right) \,,
\qquad
\Big(\CKTv\Big)(s)= k \epsilon s \,,
\qquad\Big(\CKTr\Big)(s) = k \left( \frac{\kii\epsilon^2}{2} + \frac{s^2}{2}  \right) \,, 
\end{equation}
so, for these orbits with $\sLab=0$ and no matter of the value of the curvature of the configuration space, the $s$-evolution is the following: $\CKTv$ is a linear function of $s$ and $\CKTu$ and $\CKTr$ are quadratic (further $t$ is cubic). This extends the semicubical type singularity at the cusp for the function $x(t)$, whose graph is a cycloid for the parabolic orbits in the Euclidean case.  

A point worth to remark is the following: for {\it any} $\sLab$ and in a configuration space of {\it any} curvature, $\CKTv$ is still a linear function of $\CKS_{\sLab}(s)$, which is itself the `$\sLab$-deformation' of $s$, so in some CK sense `$\CKTv$ is linear in $s$', while $\CKTu$ and $\CKTr$ are linear functions of $\CKV_{\sLab}(s)$ which the is the natural `$\sLab$-deformation' of $s^2/2$, hence they are `quadratic in $s$' in the CK sense.

\subsection{Kepler motion on a curved space $\CKspace$ as a geodesic flow in a ${\bf S}^2_{\sLab[\kii]}$}

The results obtained in the previous sections can be summed up in the form: 

\medskip

{\bf Theorem} [`curved' version of Moser, Osipov, Belbruno;  Milnor] {\sl
Consider Kepler motions of a particle in a configuration space $\CKspace$ with a metric of constant curvature $\ki$ and signature type $\kii$ (Riemannian or Lorentzian for $\kii >,<0$ respectively). 
For such a Kepler motion, there are two basic constants of motion, the energy $E$ and the angular momentum $\CKJ$. Restrict attention to the set of all Kepler motions with a fixed value of the combination $\sLab = -(2E - \ki\kii\CKJ^2)$, and look to these motions in momentum plane $\pmb{\CKP}=(\CKP_1, \CKP_2)$. On the space of all `momentum vectors' admissible for such motions, suitably completed , there is a unique metric $ds^2$, which is of signature type $\kii$, with the following properties: 
\smallskip

1) The metric has constant curvature $\sLab$.

2) The geodesics of this metric are precisely the `momentum hodographs' for the `curved' Kepler problem, and

3) The arc-lenght parameter of this metric coincides with the `curved' Levi-Civita parameter along the corresponding Kepler motion. 

\smallskip

Hence, the Kepler motion on a curved space $\CKspace$ can be seen as a geodesic flow in a ${\bf S}^2_{\sLab[\kii]}$
}

\bigskip

Two comments would help to clear up possible misunderstandings. First, the `suitable completion' mentioned in the Theorem statement is the following: When $\kii>0$, a single point at infinity $\pmb{\CKP}=\infty$ has to be added. When $\kii<0$, the completion requires to add a single point at infinity {\it and} two straight lines, which together make the light cone of the point at infinity; some detailed discussion of this issue in \cite{ConformalCompactHS02}. Second, the previous theorem refers motions with a fixed given $\sLab$ and to a metric, the `Levi-Civita one' $ds^2$ which is of constant curvature $\sLab$, in the (completed) momentum plane $\pmb{\CKP}=(\CKP_1, \CKP_2)$, turning this space into a ${\bf S}^2_{\sLab[\kii]}$. In the same momentum plane, without the completion,  we may consider another {\it flat} metric with signature type $\kii$ as well and {\it independent} of $\sLab$. The momentum hodographs of any Kepler  orbit, which are geodesics relatively to the Levi-Civita metric with the corresponding value for $\sLab$, can be alternatively seen as `cycles' relatively to the flat  metric $d\wp^2=\CKP_1^2 + \kii \CKP_2^2$. For a fixed value of $\sLab$, the cycles are precisely the ones obtained from the stereographic projection of the geodesics in ${\bf S}^2_{\sLab[\kii]}$. Of course, completion of this plane is required if we want to have an uniform description including e.g., straight lines as circles.  

\subsection{The period of the `curved' Kepler orbits and the three Kepler laws}

The major semiaxis $a=\smallonehalf(r_{per}+r_{apo})$ of the closed elliptic orbits (those with $e<1$ and real $r_{per}$ and $r_{apo}$; $\kii>0$) can be easily related to the energy and angular momentum  of the orbit. We start from (\ref{CKepOrbit}), evaluate it at $\phi=0$ (the periastron) and at $\phi=\pi/\sqrt{\kii}$ (the apoastron), and use (\ref{CKepRelsEJ_ep}) to obtain:
\begin{equation}
\CKt(r_{per})=\frac{\kii \CKJ^2}{k}\frac{1}{1+e},
\qquad
\CKt(r_{apo})=\frac{\kii \CKJ^2}{k}\frac{1}{1-e},
\end{equation}
Then expand $\CKt(2a)=\CKt(r_{per}+r_{apo})$ using the formula for addition of tangents (\cite{HeOrSa00})
\begin{equation}
\CKt(r_1+r_2)=\frac{\CKt(r_1) + \CKt(r_2)}{1-\ki\CKt(r_1)\CKt(r_2)}
\end{equation}
and the relation (\ref{CKepRelsEJ_ep}). Angular momentum dissapears and the result is
\begin{equation}
\CKt(2a) =\frac{k}{-E}
\end{equation}
hence for the closed elliptic orbits which appear in Riemannian configuration spaces of any curvature $\ki$, the energy of the motion depends {\it only} on the ellipse major semiaxis, and the relation is:
\begin{equation}
E =-\frac{k}{\CKt(2a) }
\label{CKepRelE_a}
\end{equation}
which evidently reduces in the Euclidean case $\ki=0$ to the well known result $E=-k/(2a)$. 

The period for these orbits can be also similarly obtained. Start from noticing that periastron and apoastron correspond to the two values $s=0$ and $s=2\ \quadrant_\sLab=\pi/\sqrt{\sLab}$, so for symmetry reasons the period of a closed elliptic orbit can be directly obtained as $T=2t(s=2\quadrant_{\sLab})$ in terms of formula (\ref{CKepT_s}). 
Now, as $\CKT_\sLab(\quadrant_\sLab)=\infty$, multiplication by ${\frac{\alpha+1} {\sqrt{\alpha^2-1}}}$ is ineffective, and $\CKArcT_{-\sLab}\!\left( {\frac{\alpha+1} {\sqrt{\alpha^2-1}}}\CKT_\sLab(\quadrant_\sLab)\right)=\quadrant_{\sLab}$. The same happens for the $\beta$ contribution. Thus we get the relation
\begin{equation}
\frac{e k \ki}{2 \sLab}  T = \left\{
\frac{\alpha}{\sqrt{\alpha^2-1}}
-
\frac{\beta}{\sqrt{\beta^2-1}} 
 \right\} \frac{\pi}{2\sqrt{\sLab}}\,.
 \label{CKepPeriodo1}
\end{equation}

Now if we develop this expression using (\ref{CKepDefAlphaBeta}) and (\ref{CKepRelE_a}), after some slightly tedious but straightfowward computation, the angular momentum also disappears and the period depends {\it only} on the energy, as was to  be expected, through a rather uninspiring relation:
\begin{equation}
T^2 = 
\pi^2\frac{1}{-E(1+\frac{\ki k^2}{E^2})} \frac{\sqrt{1+\frac{\ki k^2}{E^2}}-1}{\ki}
 \label{CKepPeriodo_E}
\end{equation}
and in the standard spherical case this coincides, after some algebra, with the one given in \cite{Ko92}. Again when $\ki\to0$ the limit is easily seen to be $T^2=\pi^2\frac{k^2}{-2 E^3}$. Back to the general case with  any curvature $\ki$, by using the relation (\ref{CKepRelE_a}) between energy and major semiaxis and simplifying, a much more transparent relation relating the major ellipse semiaxis to the period appears:
\begin{equation}
T^2 = \frac{\pi^2}{k} \CKs(2a)\CKv(2a) = 
\frac{2\pi^2}{k} \CKs(2a)\CKsSq(a) = \frac{4\pi^2}{k} \CKc(a)\CKsCu(a)\,,
 \label{CKepPeriodo}
\end{equation}
which reduces in the flat Euclidean configuration space to the classical result 
\begin{equation}
\left. T^2\right|_{{\bf E}^2}= \frac{\pi^2}{k}\, 2a\, \frac{(2a)^2}{2}=\frac{4\pi^2 a^3}{k}\,.
\end{equation}
For the standard hyperbolic space, the last expression in (\ref{CKepPeriodo}) was obtained also by a direct computation starting from the law of areas  by Liebmann \cite{Lieb} (in the chapter entitled {\it ``Nichteuklidische Mechanik"} of the 1905 edition). Probably it is more clear to stick to the first form, as it displays  a factor `CK linear' and a factor `CK quadratic' in the same variable $2a$, in the sense discussed before. In terms of the natural pulsation $\omega:=2\pi/T$ for the Kepler orbits, these relations can be written, for any CK space, resembling the Euclidean form of the third Kepler law known as the 1-2-3 relation:
\begin{equation}
k = \CKc(a)\, \omega^2\, \CKsCu(a)
\label{CKepPeriodo123}
\end{equation}
(the term $\CKc(a)$ becomes invisible in the flat $\ki=0$ case, where the formula reduces to $k=\omega^2 a^3$).  

In all the previous sections, most Euclidean expressions involving $E$ 
have a `curved' analogous which is formally similar when expressed in terms of $\sLab$ instead. The relations between energy and ellipse major semiaxis or period do not follow this pattern , as their `curved' forms involve directly the energy $E$, just as in the flat case. 

Hence we may formulate the three classical laws for the Kepler problem on any configuration space $\CKspace$ in the following form:

\medskip

\noindent$\bullet$ Kepler orbits are conics, with a focus at the potential center.

\noindent$\bullet$ The law of areas, in the form $\CKJ={\CKSrSq}\dot{\phi}$ is constant, always holds.

\noindent$\bullet$ For closed elliptic orbits, period $T$ (or pulsation $\omega$) and semimajor axis $a$ are related by (\ref{CKepPeriodo}, \ref{CKepPeriodo123}).

\medskip


\section{From the `curved' to the Euclidean Kepler problem}

\subsection{The Kepler problem on the Euclidean space}

Of course, setting the standard values $\ki=0, \kii=1$ for the Euclidean space, all the expressions in Section 2 reduce to the corresponding ones for the Euclidean Kepler problem. All these expressions are quite well known, and the only novelty remaining in the Euclidean case lies in  the use of a unified setting, valid for {\it any} value of the energy, by means of the parameter $\sLab$ instead of the energy $E=-\sLab/2$. Indeed most of the general `curved' expressions are very similar, mutatis mutandis, to the Euclidean ones, and the only difference is the general appearance of labelled trigonometric functions of either $u, r, y, \phi$ with its natural labels, respectively $\ki, \ki, \ki\kii, \kii$ which reduce in the Euclidean case to parabolic functions of $u=x, r, y$ and to circular functions of $\phi$. For instance, for the relation giving the $s$ evolution of $u$: 
\begin{equation}
\CKTu = \frac{k}{\sLab} \big(\CKcLab(s) - e\big) 
{\hbox{\ \  reduces in ${\bf E}^2$ to \ \ }}
u (=x) = \frac{k}{\sLab} \big(\CKcLab(s) - e\big)\,. 
\label{CKtoEKepR_UY_s}
\end{equation}
The $s$-dependence of the cartesian coordinate $y$, the radial coordinate $r$ and of the angular coordinate $\phi$ similarly reduce to: 
\begin{equation}
y(s) = J \CKsLab(s)\,,
\quad
r(s) = \frac{J^2}{k} - e\,x(s) = \frac{k}{\sLab}\, (1- e \CKcLab(s))\,,
\quad
\tan\left(\frac{\phi}{2}\right) = 
     (1+e)\sqrt{\frac{\sLab}{1-e^2}} \CKtLab\left(\frac{s}{2}\right).
\label{CKtoEKepR_tPhi_s}
\end{equation}
     
The only not directly obvious  Euclidean limiting form is the $s$-dependence of the time $t(s)$, where a cursory look to (\ref{CKepT_s}) does not disclose resemblance to any familiar Euclidean formula. The naive replacement $\ki=0$ in (\ref{CKepT_s}) gives an indeterminate expression of the type $0/0$.  The reason of course comes from the $\ki$ dependence of the integrand in (\ref{CKIntT}); when $\ki=0$ the integration is immediate and does not call for a decomposition into simple fractions. A more careful analysis of the situation involves dealing with the $\ki\to0$ limit. This is an exercise which requires a careful consideration of square root determinations and signs, but is otherwise straightforward, so details will not be given here. The result is the one one should expect: in terms of a power expansion in $\ki$, (\ref{CKepT_s}) appears as: 
\begin{equation}
t(s) =  k \frac{s - e \CKsLab(s)}{\sLab} + O[\ki]
\label{CKepT_sLimit}
\end{equation}
which directly recalls the Kepler equation. The precise connection will be discussed in the next section.

\subsection{The Levi-Civita parameter $s$ versus the eccentric anomaly}

Traditionally, the parameter used to describe the evolution along Euclidean Kepler orbits, the {\it eccentric anomaly}, denoted $\xi$, can be introduced for either type of orbits (here we follow \cite{Wi41, ArKoNe93}) as:
\def\MSac{\hskip-7pt}
\begin{equation}
\begin{array}{lllll}
x=&\MSac a(\cos\xi -e), & y=&\MSac a\sqrt{1-e^2} \sin\xi & \hbox{when $E<0; \quad a=\frac{k}{-2E}$}\\[4pt]
x=&\MSac\smallonehalf (p-\xi^2), & y=&\MSac \sqrt{p} \, \xi, & \hbox{when $E=0$}\\[4pt]
x=&\MSac a(\cosh\xi -e), & y=&\MSac a\sqrt{1-e^2} \sinh\xi & \hbox{when $E>0; \quad a=\frac{k}{2E}$}\,.
\end{array}
	\label{EKepEccAnomEPH}
\end{equation}
The parameter $\xi$ is defined so that, for any negative energy, a complete revolution along the orbit corresponds to an increase of $\xi$ from $0$ to $2\pi$. When the energy is positive, $\xi$ takes all real values, and its scaling is chosen in analogy with the case of elliptic orbits. By replacing $x(s), y(s)$ as taken from (\ref{CKtoEKepR_UY_s}, \ref{CKtoEKepR_tPhi_s}) and comparing, one finds that $s$ and $\xi$ are related by simple scalings:
\begin{equation}
\begin{array}{llllll}
\xi =&\MSac  \sqrt{\sLab} s, 
& \cos\xi= \CKcLab(s), & \sin\xi= \sqrt{\sLab}\CKsLab(s)& 
  \hbox{when $E<0$}\,,\\[4pt]
\xi =&\MSac  \sqrt{k} s, \quad & & & \hbox{when $E=0$}\,,\\[4pt]
\xi =&\MSac  \sqrt{-\sLab} s,
& \cosh\xi= \CKcLab(s), & \sinh\xi = \sqrt{-\sLab}\CKsLab(s) & 
  \hbox{when $E>0$}\,,\\[4pt]
\end{array}
	\label{EKepEccAnomRelS}
	\end{equation}
so that along any orbit, the eccentric anomaly $\xi$ is {\it proportional} to $s$, but the coefficient  depends on the value of the orbit energy, and is not continuous at $E=0$ (even dimensionally this `parabolic' eccentric anomaly differs from the `elliptic' and `hyperbolic' ones and furthermore, different authors define it differently, compare \cite{LandauI}). 
All the `universal' equations in the previous section involving $s$ can be rewritten in terms of $\xi$, but this conceals (specially when $E=0$) the underlying similarity. From our present viewpoint, this can be seen as an unwanted artifact of the normalization imposed to the eccentric anomaly for the general elliptic and hyperbolic orbits, in a way which precludes a non-trivial limit for $\xi$ when $\sLab\to0$ (as clear in (\ref{EKepEccAnomRelS})). For instance, the Euclidean  equations for $t(s)$ (\ref{CKepT_sLimit} with $\ki=0$) and $\phi(s)$ (\ref{CKtoEKepR_tPhi_s}) reduce in three three types of orbits with $E<,=, > 0$ respectively to:
\begin{equation}
t = \left\{ 
\begin{array}{l}
\displaystyle\sqrt{\frac{a^3}{k}} (\xi - e \sin\xi), \\[4pt]
\displaystyle\sqrt{\frac{p^2}{4k}} (\xi +\frac{\xi^3}{3p}), \\[4pt]
\displaystyle\sqrt{\frac{a^3}{k}} (e \sinh\xi-\xi), \\[4pt]
\end{array}
\right.
\qquad 
\displaystyle\tan\frac{\phi}{2}=\left\{ 
\begin{array}{l}
\displaystyle\sqrt{\frac{1+e}{1-e}} \tan\frac{\xi}{2}, \\[8pt]
\displaystyle\frac{\xi}{p}, \\[4pt]
\displaystyle\sqrt{\frac{e+1}{e-1}} \tanh\frac{\xi}{2}, \\[4pt]
\end{array}
\right. 
\label{EKepTPhi_EccAn}
\end{equation}

Summing up: the facility of using the unscaled circular or trigonometric functions of the eccentric anomalies in the description of {\it all orbits with non-zero energy}, as made evident in (\ref{EKepEccAnomEPH}, \ref{EKepTPhi_EccAn}), comes at the price of losing the manifest universality of the approach using $s$ as a basic variable instead. In this approach, the functions of $s$ which appear naturally have to be the scaled associated functions $\CKcLab(s), \CKsLab(s)$ whose CK label is $\sLab=-2E$ but this description remains smooth when $E\to0$. 

\section*{Acknowledgements}
Support of projects  MTM-2005-09183 and VA-013C05 is acknowledged. This work has been also supported by a {\it Beca de Colaboraci\'on} granted by the {Ministerio de Educaci\'on y Ciencia} to L.G.G. during the academic course 2006--2007 which is also acknowledged.    

\def\otherrefs#1{}

\small 


\end{document}